# Weak-lensing Analysis of the Complex Cluster Merger A746 with Subaru/Hyper Suprime-Cam

K. HyeongHan[1], H. Cho[1,2], M. James Jee[1,3], D. Wittman[3], S. Cha[1], W. Lee[1], K. Finner[4], K. Rajpurohit[5], M. Brüggen[6], W. Forman[5], C. Jones[5], R. van Weeren[7], A. Botteon[8], L. Lovisari[5,9], A. Stroe[5], P. Domínguez-Fernández[5], E. O'Sullivan[5], and J. Vrtilek[5]

[1] Department of Astronomy, Yonsei University, 50 Yonsei-ro, Seoul 03722, Republic of Korea; ahrtears54@yonsei.ac.kr, mkjee@yonsei.ac.kr
[2] Center for Galaxy Evolution Research, Yonsei University, 50 Yonsei-ro, Seoul 03722, Republic of Korea
[3] Department of Physics and Astronomy, University of California Davis, Davis, CA 95616 USA
[4] IPAC, California Institute of Technology, 1200 E California Blvd., Pasadena, CA 91125, USA
[5] Center for Astrophysics | Harvard & Smithsonian, 60 Garden St., Cambridge, MA 02138, USA
[6] Universität Hamburg, Hamburger Sternwarte, Gojenbergsweg 112, D-21029, Hamburg, Germany
[7] Leiden Observatory, Leiden University, P.O. Box 9513, 2300 RA Leiden, The Netherlands
[8] INAF–IRA, Via P. Gobetti 101, I-40129 Bologna, Italy
[9] INAF-IASF Milano, Via A. Corti 12, I-20133 Milano, Italy
*Received 2023 November 1; revised 2023 December 20; accepted 2024 January 5; published 2024 February 12*

## Abstract

The galaxy cluster A746 ($z = 0.214$), featuring a double radio relic system, two isolated radio relics, a possible radio halo, disturbed V-shaped X-ray emission, and intricate galaxy distributions, is a unique and complex merging system. We present a weak-lensing analysis of A746 based on wide-field imaging data from Subaru/Hyper Suprime-Cam observations. The mass distribution is characterized by a main peak, which coincides with the center of the X-ray emission. At this main peak, we detect two extensions toward the north and west tracing the cluster galaxy and X-ray distributions. Despite the ongoing merger, our estimate of the A746 global mass $M_{500} = 4.4 \pm 1.0 \times 10^{14} M_\odot$ is consistent with the previous results from Sunyaev-Zel'dovich and X-ray observations. We conclude that reconciling the distributions of mass, galaxies, and intracluster medium with the double radio relic system and other radio features remains challenging.

*Unified Astronomy Thesaurus concepts:* Abell clusters (9); Galaxy clusters (584); Weak gravitational lensing (1797)

*Supporting material:* machine-readable table

## 1. Introduction

At the highest primordial density peaks, galaxy clusters form and grow in size and mass through the hierarchical structure formation paradigm by aggregating dark matter, gas, and galaxies. Among the cluster evolutionary channels, mergers are the most cataclysmic events, releasing kinetic energies up to ∼$10^{64}$ erg (Markevitch et al. 1999; Ricker & Sarazin 2001). During the collision, a small fraction of the energy is dissipated into the intracluster medium (ICM) through weak shocks ($\mathcal{M} \lesssim 4$) and turbulence, which, on megaparsec scales, accelerate electrons to relativistic speeds and amplify magnetic fields. The cluster-wide diffuse synchrotron radiation produced by these processes is broadly classified into two categories, radio relics and halos, depending on the observed properties such as polarization, morphologies, and locations.

Radio relics are polarized, elongated synchrotron sources typically found in the periphery of cluster mergers (see Brunetti & Jones 2014; van Weeren et al. 2019, for a review and references therein). Their morphological traits are believed to trace merger shock waves generated during the core passage, providing us with insights into the merger history and configuration (i.e., merger axis, time since collision, collision velocity, impact parameter, etc.). On the other hand, radio halos are unpolarized, steep-spectrum ($\alpha > 1$) synchrotron radiation that are spatially correlated with the ICM distribution (e.g., Cassano et al. 2010; Cuciti et al. 2015), extending over the central region of the cluster. They are often found in massive merging galaxy clusters ($M_{500} \gtrsim 5 \times 10^{14} M_\odot$), suggesting that the merger-induced turbulence re-accelerates energetic plasma particles to the relativistic scale (e.g., Brunetti & Jones 2014, for a review).

Double radio relic systems are a rare subclass where two relics are observed on opposite sides, bracketing the merger (Bonafede et al. 2009; van Weeren et al. 2010). In particular, a symmetric[10] double radio relic cluster exhibits a bimodal distribution of galaxies and dark matter, which aligns with the vector connecting the two relics (e.g., Jee et al. 2015; Finner et al. 2021; Kim et al. 2021; Cho et al. 2022; Lee et al. 2022). Because the two relics trace the two merger shocks originating from the same collision and traveling in opposite directions, they generally resolve merger ambiguity and provide tight constraints on the merger scenario.

A746 at $z = 0.214$ is an exceptional cluster merger featuring a symmetric east–west double radio relic system (van Weeren et al. 2011; Botteon et al. 2022) with an intricate, non-bimodal galaxy distribution (Golovich et al. 2019a, 2019b). Also, the cluster hosts a highly disturbed, V-shaped distribution of ICM (Rajpurohit et al. 2023), which overlaps the smaller relic. In addition to the double radio relic system, there are two more fainter and isolated radio relics on the northern and eastern peripheries. At present, it is challenging to construct a coherent merging scenario that can account for the symmetric double radio relic system, the complex galaxy distribution, the disturbed X-ray morphology, and the presence of the two isolated diffuse radio emissions.

---

[10] Here we define symmetric radio relics as the system where the two normal vectors to the double radio relics are close to antiparallel.

In this study, as a crucial step toward understanding A746, we perform weak lensing (WL) analysis using Subaru/Hyper Suprime-Cam (HSC; Furusawa et al. 2018; Kawanomoto et al. 2018; Komiyama et al. 2018; Miyazaki et al. 2018) observations and re-examine the cluster galaxy distribution with MMT/Hectospec spectroscopic data. Although galaxies are biased tracers of the underlying dark matter, they serve as useful proxies for the identification of merging substructures (e.g., Guennou et al. 2014; Dawson et al. 2015; Golovich et al. 2017). The previous study on the galaxy distribution of A746 is largely based on 66 members observed with Keck/DEIMOS (Golovich et al. 2019a). The presence of the f UMa star ($V = 4.48$; Ducati et al. 2001) limited the ability to identify cluster members on the western periphery. In the current study, we increase the sample size of the spectroscopic members with MMT/Hectospec data and supplement the cluster member catalog with photometric members after carefully subtracting the f UMa star. Together with the galaxy distribution, we use our WL result to resolve merging substructures. WL allows us to robustly quantify the cluster mass, which can be biased in other measurements (such as Sunyaev-Zel'dovich (SZ), X-ray, and velocity dispersion) due to the ongoing merger. Finally, WL provides insights into the dark matter substructures (e.g., Clowe et al. 2006; Jee & Tyson 2009; Okabe et al. 2011; Ragozzine et al. 2012; HyeongHan et al. 2020; Cho et al. 2022; Finner et al. 2023), which are quintessential components in understanding the system since they govern the dynamics of the cluster merger.

This paper is organized as follows. In Section 2, we describe our observing design and data reduction process of Subaru/HSC and MMT/Hectospec observations. The theoretical background of the WL analysis and our shear estimation is described in Section 3. We present the reconstructed mass map in Section 4. The mass estimation from the WL and spectroscopy analysis is presented in Section 5. We compare the radio relic luminosity to the WL mass along with other systems in Section 6. In Section 7, we suggest a merger scenario based on the mass and galaxy distribution with the X-ray and radio observations. We summarize our results in Section 8.

We adopt the cosmological parameters of $H_0 = 70$ km s$^{-1}$ Mpc$^{-1}$, $\Omega_M = 0.3$, and $\Omega_\Lambda = 0.3$ under a flat $\Lambda$ cold dark matter ($\Lambda$CDM) cosmology throughout this paper. $R_\Delta$ represents the radius where the average density becomes $\Delta$ times the critical density of the universe. $M_\Delta$ is the total mass within $R_\Delta$. At the redshift of A746 ($z = 0.214$), the plate scale is 3.478 kpc arcsec$^{-1}$.

## 2. Observations

### 2.1. Subaru Imaging

We observed A746 with the Subaru HSC on 2023 January 16, 20, and 22 (PI: H. Cho). The total integrations are 3480 and 4320 s in the HSC-$g$ and HSC-$r2$ bands, respectively, which cover $\sim$2.4 deg$^2$. We employed a strategy of dithering and rotation between exposures in the observation plan. This method effectively reduces the occurrence of CCD blooming artifacts and diffraction spikes caused by bright stars, thereby mitigating their influence on the WL analysis of faint source galaxies. The median on-site seeing conditions for $g$ and $r2$ bands were FWHM=$1.''26$ and $0.''75$, respectively.

The very bright star f UMa with $V = 4.48$ mag is located near the northwestern radio relic $7.'4$ ($\sim$1.6 Mpc) away from the X-ray center of A746 (see Figure 1). The emission from the extended point-spread function (PSF) of this star results in a luminous halo, limiting our ability to detect galaxies. The detection of galaxies can be further complicated due to the optical ghosts caused by unwanted reflections on various surfaces within the HSC optical path, particularly for stars brighter than $\sim$7 mag (Aihara et al. 2022). In general, positioning such an extremely bright star outside the field of view of the instrument is advised to minimize the aforementioned issues. Nevertheless, we opted to configure the telescope pointings at the approximate center of f UMa. This makes the emission from the halo and reflection patterns appear azimuthally symmetric, which helps us to characterize and subtract them (Appendix A). Since A746 is located near f UMa, this pointing scheme also maximizes the clustocentric radius, at which we can complete a circle.

Construction of the calibration data and single-frame calibration (overscan/bias/dark subtraction, flat fielding, astrometric correction, etc.) were carried out with the LSST Science Pipelines stack v22_0_0 (Bosch et al. 2018, 2019). Throughout the calibration process, the pipeline adopts the Simple Imaging Polynomial (SIP) convention to represent distortions in the calibrated FITS images. Following the HSC techniques of Finner et al. (2023) and HyeongHan et al. (2023), we converted SIP to TPV[11] using the `sip_tpv` code[12] (Shupe et al. 2012) in order to guarantee compatibility with the `SWarp` software[13] (Bertin et al. 2002), which is used in the later stacking process.

### 2.2. Photometry

We coadded our final mosaic image using the `SWarp` clipped mean stacking method (Gruen et al. 2014) in both filters. To maximize the number of source detections, we created a detection image by combining the $g$ and $r2$ mosaic images. We ran `SExtractor`[14] (Bertin & Arnouts 1996) in dual-image mode. We utilized a weight image generated by `SWarp` to create the detection and root-mean-square error images for photometry (e.g., Finner et al. 2017; HyeongHan et al. 2020; Finner et al. 2021; HyeongHan et al. 2023). During the analysis, we employed `MAG_ISO` for color estimation and `MAG_AUTO` for luminosity. We performed photometric calibration relative to the Sloan Digital Sky Survey (SDSS DR16; Ahumada et al. 2020) with the crossmatched objects in the same field.

### 2.3. MMT/Hectospec Spectroscopy

Spectroscopic observations (PI: K. Finner) were conducted with Hectospec, a multi-object fiber spectrograph (Fabricant et al. 2005) at the MMT Observatory. We selected spectroscopic targets for A746 within the 1° diameter field of view of the Hectospec instrument based on archival imaging from

---
[11] https://fits.gsfc.nasa.gov/registry/tpvwcs/tpv.html
[12] https://github.com/stargaser/sip_tpv
[13] http://www.astromatic.net/software/swarp
[14] https://github.com/astromatic/sextractor

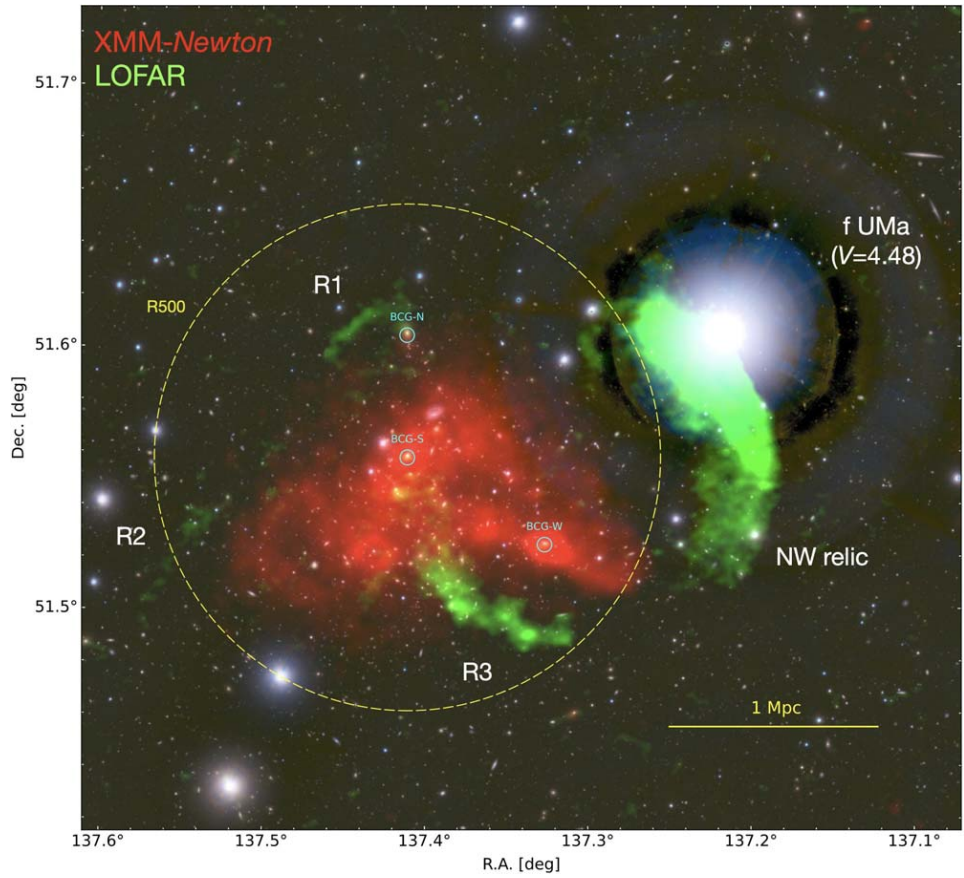

**Figure 1.** Multiwavelength view of A746. The background pseudo-color composite image is created by using $r2$, $g + r2$, and $g$ filters for the red, green, and blue channels, respectively. The X-ray (XMM-Newton; Rajpurohit et al. 2023) and radio (LOFAR; Botteon et al. 2022; Rajpurohit et al. 2023) emissions are color coded in red and green, respectively. The confirmed radio relics (NW relic and R1), candidate relics (R2 and R3) reported by Rajpurohit et al. (2023) and the bright star (f UMa) are annotated. The three BCGs (BCG-N, BCG-S, and BCG-W) are marked with cyan circles. The yellow dashed circle indicates the $R_{500}$ radius centered at BCG-S.

**Table 1**
Spectroscopic Redshift Catalog of Galaxies in the A746 Field

| R.A. (J2000) (deg) (1) | Decl. (J2000) (deg) (2) | $cz$ (km s$^{-1}$) (3) | $z$ (4) | $z_{\mathrm{err}}$ (5) | $R_{\mathrm{XC}}$ (6) | Catalog Source (7) |
|---|---|---|---|---|---|---|
| 137.55600833 | 51.86678694 | $55{,}551 \pm 23$ | 0.185298 | $7.78747 \times 10^{-5}$ | 18.8 | MMT |
| 137.23538750 | 51.83154306 | $55{,}178 \pm 18$ | 0.184056 | $5.97792 \times 10^{-5}$ | 20.2 | MMT |
| 137.56300417 | 51.91351306 | $69{,}300 \pm 42$ | 0.231163 | $1.39759 \times 10^{-4}$ | 8.9 | MMT |
| 137.39141667 | 51.58300778 | $64{,}344 \pm 34$ | 0.214629 | $1.12756 \times 10^{-4}$ | 11.8 | MMT |
| 137.48850417 | 51.73325361 | $74{,}839 \pm 60$ | 0.249637 | $1.99713 \times 10^{-4}$ | 5.2 | MMT |
| 137.55725417 | 51.97468944 | $69{,}342 \pm 22$ | 0.231302 | $7.49948 \times 10^{-5}$ | 15.4 | MMT |
| 137.23352917 | 51.98109444 | $78{,}774 \pm 20$ | 0.262762 | $6.52144 \times 10^{-5}$ | 17.8 | MMT |
| 137.87894583 | 51.58250417 | $64{,}434 \pm 33$ | 0.214931 | $1.10062 \times 10^{-4}$ | 12.5 | MMT |
| 137.62995417 | 51.54395667 | $63{,}968 \pm 16$ | 0.213376 | $5.25009 \times 10^{-5}$ | 21.1 | MMT |
| 137.86744583 | 51.82847972 | $78{,}584 \pm 51$ | 0.262129 | $1.69225 \times 10^{-4}$ | 8.1 | MMT |

**Notes.** Table 1 is published in its entirety in machine-readable format. A portion is shown here for guidance regarding its form and content. The columns are as follows: (1) R.A. and (2) decl. in degrees (J2000.0), (3) redshift velocity and measurement uncertainty in kilometers per second, (4) redshift and (5) uncertainty, and (6) cross-correlation score $r$-value obtained from RVSAO (only available for the MMT catalog). Column (7) provides the references for the catalog data. The codes MMT, G19, SDSS, and NED refer to MMT/Hectospec observations, Golovich et al. (2019a), SDSS DR18, and NED, respectively.

(This table is available in its entirety in machine-readable form.)

Subaru/Suprime-Cam and SDSS and photometric redshift data. Primary cluster member candidates were selected if they were found within a range of $|\Delta(g - r)| < 0.05$ mag of a linear fit to the red sequence. Also, candidates redder or bluer than the red sequence were included if their photometric redshifts were within $\pm 0.05$ from the cluster redshift.

The MMT/Hectospec observations provided a total of 244 spectra (Table 1). The data were reduced with the HSRED v2.1 pipeline.[15] We derived redshifts by cross-correlating each of

---
[15] https://github.com/MMTObservatory/hsred

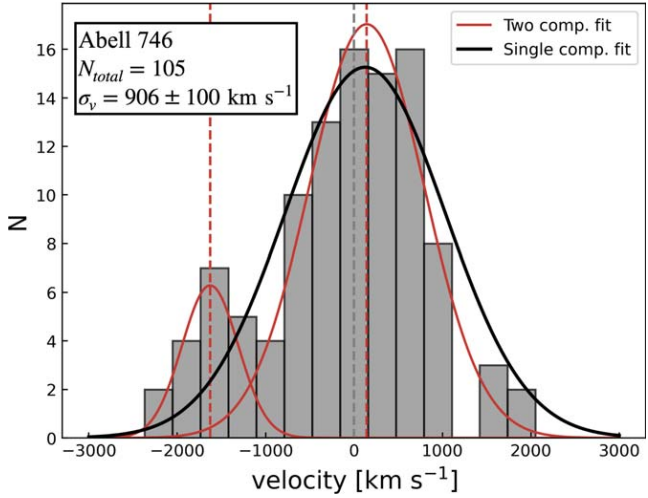

**Figure 2.** Relative velocity distribution of galaxies in A746. The black solid curve represents the best-fit Gaussian. The red dashed lines indicate the mean values of the two-component Gaussian fit. The relative velocity of BCG-S is marked by the gray dashed line. From the statistical inference (see the text for more details), the single Gaussian fit is marginally preferred.

244 spectra with a set of template spectra using the xcsao task of the IRAF package RVSAO (Kurtz & Mink 1998). Based on the criteria discussed in Kurtz & Mink (1998), we assessed the reliability of the redshift measurement using a cross-correlation score, denoted as the $r$-value (Tonry & Davis 1979; $R_{XC}$), which needed to exceed 4. We successfully obtained reliable redshifts for 229 objects.

### 2.4. Galaxy Cluster Membership Determination

In addition to our MMT/Hectospec data, we compiled archival spectroscopic data from Keck/DEIMOS (Golovich et al. 2019a), SDSS DR18 (Almeida et al. 2023), and NED. To determine the cluster membership, we first selected galaxies within the projected virial radius ($\sim 10'$) from the southern brightest cluster galaxy (BCG-S in Figure 1). Then, we iterated $3\sigma$ clipping for the galaxies having a radial velocity offset less than $\pm 3000$ km s$^{-1}$ from the literature mean cluster redshift ($z = 0.215$; Golovich et al. 2019a, 2019b) until it converged. This process leaves us with 105 member galaxies with a mean redshift of $z = 0.214$.

Figure 2 shows the velocity distribution of the member galaxies. When we fit a single Gaussian (black), the best-fit velocity dispersion is $\sigma_v = 906 \pm 100$ km s$^{-1}$, consistent with that in Golovich et al. (2019a), $\sigma_v = 1094 \pm 95$ km s$^{-1}$. Considering the possibility of a substructure corresponding to a small peak at $\Delta v \sim -2000$ km s$^{-1}$, we also performed a two-component Gaussian fit (red). Although the residual is smaller in the two-component case, our tests with Bayesian information criterion (BIC) and Akaike information criterion (AIC) show that the single-component fit is marginally favored [BIC$_1$ − BIC$_2$ = −1.3 and exp((AIC$_1$ − AIC$_2$)/2) = 0.3]. In addition, we find no spatial clustering on the projected plane for the galaxies belonging to the secondary peak.

In order to explore the spatial distribution of the galaxies, we used the cluster member catalog and supplemented it with photometric members. We began by identifying the locus of the red-sequence galaxies in the color–magnitude diagram using the 105 spectroscopic members. This process involved applying $2\sigma$ clipping and performing a linear fit. Then, we selected photometric members with $(g - r2) < 1\sigma$. The faint end is set to $r2 \sim 21$ mag, which is 4 magnitudes fainter than the brightest spectroscopic member (see the cyan box in Figure 3).

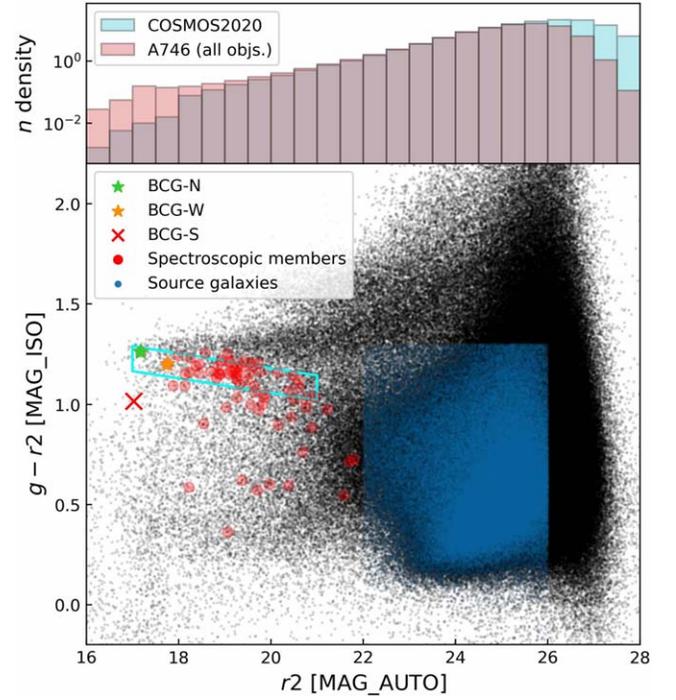

**Figure 3.** Luminosity function (top) and color–magnitude relation (bottom) for A746. BCG-N and -W are marked by a star symbol with different colors (see Figure 1). The red cross refers to BCG-S, which is blended with a blue background galaxy (see Appendix B and Figure 1). The spectroscopic members are marked in red circles. The blue dots indicate source galaxies for the WL analysis. The cyan box represents the selection window for the photometric member selection (see Section 2.4).

## 3. Weak-lensing Analysis

### 3.1. Basic Weak-lensing Theory

Weak-lensing (WL) measures the effects of the tidal deflection of the light path induced by inhomogeneities in the matter density field along the line-of-sight (LOS) direction. The shapes of the background objects are modified on the image plane (for reviews, see Schneider 2005; Mandelbaum 2018, and references therein), according to the Jacobian matrix $A$:

$$A = \begin{bmatrix} 1 - \kappa - \gamma_1 & -\gamma_2 \\ -\gamma_2 & 1 - \kappa + \gamma_1 \end{bmatrix}, \quad (1)$$

where $\gamma_{1(2)}$ and $\kappa$ are the shear and convergence, respectively. The shear causes an anisotropic shape distortion, which is often expressed by a complex notation $\gamma = \gamma_1 + i\gamma_2$. The convergence $\kappa$ refers to the projected mass density in units of the critical surface density:

$$\Sigma_c = \frac{c^2 D_s}{4\pi G D_l D_{ls}}, \quad (2)$$

where $c$ is the speed of light, $G$ is the gravitational constant, $D_l$ is the angular diameter distance to the lens, $D_{ls}$ is the angular diameter distance from the lens to the source, and $D_s$ is the angular diameter distance to the source. The effects of $\gamma$ and $\kappa$ are inseparable, and in reality, we only observe the reduced shear $g = \gamma/(1 - \kappa)$.

### 3.2. Point-spread Function Modeling and Shape Measurement

Our shape measurement pipeline is comprised mainly of two parts: point-spread function (PSF) modeling and forward-model fitting of galaxy shapes. We performed a principal component analysis (PCA; Jee et al. 2007; Jee & Tyson 2011) of the observed stars for each resampled frame to capture CCD-to-CCD variations in the size and ellipticity of PSFs. From the PCA procedure, we obtained the covariance matrix and calculated the eigenvectors and eigenvalues. We chose 21 principal components (PCs) that are responsible for the 21 highest variances. To account for the spatial variation across the CCDs, we fitted third-order polynomials to the coefficients of the PCs. At the location of an object in the final mosaic image, we modeled the individual PSFs for all associated resampled frames and then stacked them to create the PSF for the mosaic image.

We measured the shapes of galaxies by fitting a PSF-convolved elliptical Gaussian function to a square postage-stamp cutout. These shapes embody biases that dilute the lensing signal. The imperfect modeling of a real galaxy introduces *model bias*, the nonlinearity between pixel noise and shear causes *noise bias*, and the finite size of the cutout image induces *truncation bias*. The blending effect that contaminates a large fraction of detected objects is another important contributor to the bias (e.g., Dawson et al. 2016). We derived the correction factor by iteratively running our WL pipeline for a set of simulated images using the GalSim (Rowe et al. 2015) package that resemble the properties of the observed sources such as crowdedness, ellipticity distribution, luminosity function, Sérsic index, etc. The simulation also includes the effect of spatially varying PSFs. By iteratively comparing the input (true) shear of the simulations with the measured shear, we derived a multiplicative bias of $m = 1.15$ (Jee et al. 2016) after applying the same source selection criteria described in Section 3.3. We found that the additive bias is negligible. Readers are referred to Jee et al. (2013) for further details on the shear calibration, and Finner et al. (2017, 2020) and Cho et al. (2022) for the detailed implementation of the shear measurement.

### 3.3. Source Selection and Redshift Estimation

Ideally, one can select sources based on high-fidelity photometric redshift information. The limited number of filters, however, makes this approach impractical in the current study. Instead, we relied on the traditional method of utilizing the color–magnitude relation (Figure 3). Considering the redshift of A746 and the location of the so-called 4000 Å break, the cluster red-sequence galaxies are readily identified with their $g - r2$ colors.

To determine the magnitude interval of the source population, we compared the number density of our source catalog to that of the COSMOS2020 field (Weaver et al. 2022). We chose the magnitude lower limit as $r2 > 22$, below which the excess due to the presence of the cluster members is found. Objects fainter than $r2 = 26$ mag were discarded since they are dominated by the low signal-to-noise ratio ($<10$) galaxies, whose WL shape measurements are unstable. We used the red-sequence color to determine the red end of the source color ($g - r2 < 1.3$).

In addition to the above photometric criteria, we imposed the following requirements on shapes (e.g., Jee et al. 2013): (1) the elliptical Gaussian fitting should be successful (STATUS = 1), (2) the semiminor axis should be larger than 0.4 pixels, and (3) the ellipticity measurement error should be less than 0.3. We then rejected sources affected by the diffraction patterns of bright stars after visual inspection. The objects that satisfy the above conditions are shown in blue in the color–magnitude diagram (Figure 3), and the resulting source density is 24 arcmin$^{-2}$.

The WL signal is proportional to the angular diameter distance ratio $\beta = D_{ls}/D_s$. We calculate its effective mean as follows:

$$\langle \beta \rangle = \left\langle \max\left(0, \frac{D_{ls}}{D_s}\right) \right\rangle. \quad (3)$$

We utilized the COSMOS2020 photometric redshift catalog (Weaver et al. 2022) as a control field to infer the source redshift distribution in the A746 field.

We applied the photometric selection as in the source selection scheme to the COSMOS2020 catalog and measured the number density ratio per magnitude bin between the COSMOS2020 and A746 fields. The ratio is used to account for the difference in depth between the two fields.

We obtained $\langle \beta \rangle = 0.63$, which corresponds to an effective redshift of $z_{\rm eff} = 0.66$. Using a single representative value for the source population introduces bias because the relation between the lensing efficiency $\beta$ and redshift is nonlinear. To address the redshift distribution of the sources, we applied a first-order correction to the reduced shear $g' = [1 + \kappa(\langle \beta^2 \rangle/\langle \beta \rangle^2 - 1)]g$ (Seitz & Schneider 1997) where $\langle \beta^2 \rangle = 0.47$.

### 4. Projected Mass Reconstruction

The Fourier-inversion method (Kaiser & Squires 1993) has been widely used in the community to reconstruct the convergence ($\kappa$) map from the shear field. We obtained the convergence field through the following convolution:

$$\kappa(\theta) = \frac{1}{\pi} \int D^*(\theta - \theta') \gamma(\theta') d^2\theta', \quad (4)$$

where

$$D^*(\theta) = \frac{\theta_1^2 - \theta_2^2 - 2i\theta_1\theta_2}{|\theta|^4} \quad (5)$$

is the convolution kernel. We bootstrapped the shear catalog to generate 2000 realizations, which were combined to create a convergence rms (noise) map. A significance map was obtained by dividing the convergence map by the rms map.

Figure 4 shows the mass reconstruction (white contours) of A746. The main mass peak, despite the lack of a distinct BCG at its location, coincides with the approximate centers of the X-ray emission and cluster galaxy distribution. The centroid uncertainties of the mass peaks were estimated by applying a Gaussian prior to the peak distributions from the bootstrapped convergence maps (magenta contours in the right panel). At the main mass peak, the mass distribution is elongated in both east–west and north–south directions. These complex features consistently appear when we repeat our mass reconstructions with different algorithms, such as FIATMAP (Fischer & Tyson 1997; Wittman et al. 2006) and MARS (Cha & Jee 2022). The galaxy distribution in A746 follows this complex mass distribution.

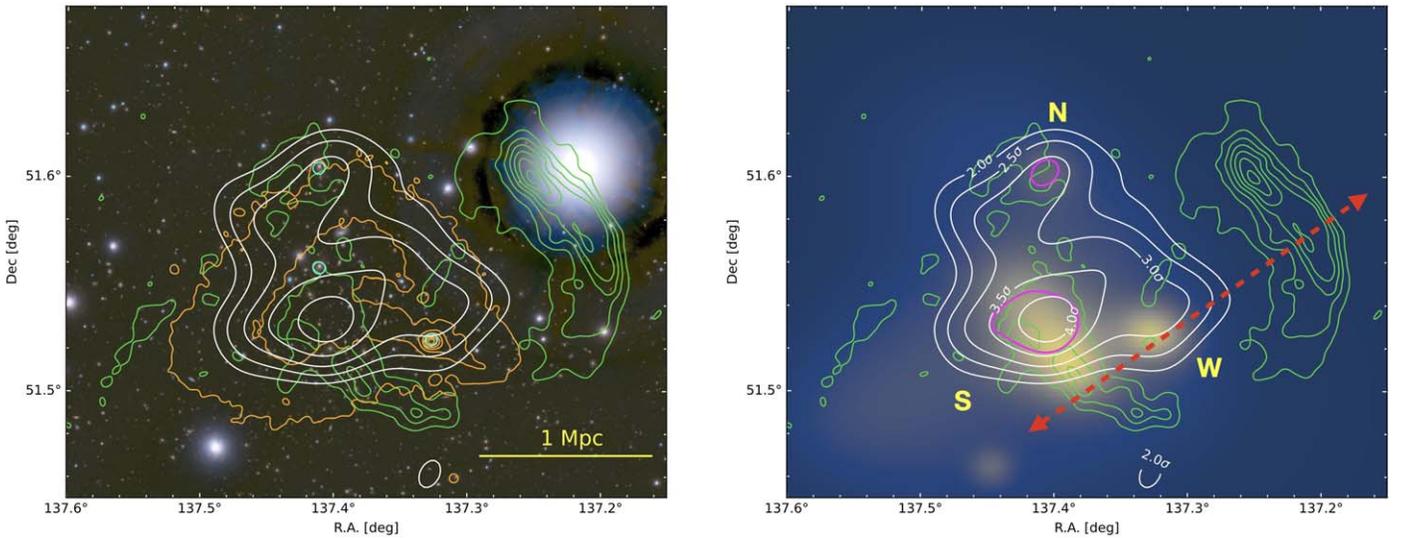

**Figure 4.** Projected mass distribution of A746. For both panels, the white contours represent the significance of the mass reconstruction. The lowest contour indicates the $2\sigma$ level and the contour spacing is $0.5\sigma$. Left: The XMM-Newton observation is shown in orange, and the LOFAR radio observation is in green (Rajpurohit et al. 2023). The disturbed features of the X-ray emission and mass map are in the east–west and north–south directions. The cyan circles indicate the BCGs for each subhalo. Right: The background is the galaxy number density map adaptively smoothed by the csmooth script in the ciao-4.12 package with a minimum (maximum) significance of $1.5\sigma$ ($3\sigma$). The minimum smoothing scale corresponds to $\sim 10''$. The magenta contours represent the $1\sigma$ mass centroid uncertainty from the bootstrapping analysis. The mass centroids are in spatial agreement with the X-ray and galaxy distribution. The red arrows indicate the hypothesized merger axis defined by the double radio relics (NW and R3).

The presence of the bright star poses a substantial hindrance to both the identification of galaxies and the weak-lensing analysis for the $r \lesssim 4'$ region from it. The luminous stellar halo prevents reliable detection of small, faint galaxies. Also, artifacts due to the reflection ring can lead to the false detection of highly sheared background galaxies around the star. Given the proximity of this bright star from A746, it is imperative to ensure that our results are not significantly influenced by its artifacts. We report that we have found no hints of mass or galaxy overdensity near the f UMa star after we carefully subtracted this extremely bright star and then searched for astronomical sources in its vicinity. See Appendix A for details on the subtraction method.

## 5. Mass Estimation

Given the complex distributions of the mass, galaxy, and ICM, in addition to the multiple radio relics, A746 might be undergoing multiple mergers. However, our mass reconstruction does not resolve the cluster mass distribution into a combination of distinct subhalos, making it infeasible to characterize the cluster mass based on multiple profile fitting (e.g., Jee et al. 2016; Finner et al. 2017; HyeongHan et al. 2020; Finner et al. 2021; Cho et al. 2022; Finner et al. 2023). Even when we forced multi-halo fitting, the result did not converge. In this study, we thus limit ourselves to estimating the global mass of A746.

We choose the BCG nearest to the mass peak (BCG-S in Figure 1) as the cluster center. The BCG is $\sim 2'$ offset from the mass and X-ray centers. We constructed an azimuthally averaged tangential shear profile. The reduced tangential ($g_+$) component is calculated by the following equation:

$$g_+ = -g_1 \cos 2\phi - g_2 \sin 2\phi, \quad (6)$$

where $g_1$ and $g_2$ are the components of the reduced shear, and $\phi$ is the position angle with respect to the cluster center. Figure 5 shows the resulting tangential shear profile. By rotating the galaxy position angle by $\pi/4$, we obtain the cross-shear (B-mode) component ($g_\times$). It provides a diagnostic test for the residual systematic error, which is consistent with zero over wide radii ($R \leqslant 20'$) in our analysis. A large number of source galaxies in each bin ($\sim 10^3$) secures the statistical stability over the masked stellar region, which is $\sim 8'$ away from the center.

Mass estimation is performed by fitting 1D density models to the reduced shear profile. We masked the profile at $r \lesssim 5'$ since the lensing signal suffers (1) cluster member contamination, (2) the centroid bias, (3) nonlinearity of the shear response, and (4) the baryonic effect. The $r < R_{500}$ ($5'.8$ or $\sim 1.2$ Mpc) region encloses the three subhalos (N, S, and W) where the cluster potential is expected to deviate from the assumed profile, potentially leading to non-negligible mass bias (Lee et al. 2023).

The singular isothermal sphere (SIS) model yields a velocity dispersion of $\sigma_v = 902 \pm 77 \text{ km s}^{-1}$ with a reduced $\chi^2$ of 1.0. This value is in excellent agreement with the direct measurement ($\sigma_v = 906 \pm 100 \text{ km s}^{-1}$, Section 2.4). The Navarro–Frenk–White (NFW; Navarro et al. 1996, 1997) halo fitting with the mass–concentration relation of Ishiyama et al. (2021) provides a mass of $M_{200} = 6.3 \pm 1.5 \times 10^{14} M_\odot$ with a reduced $\chi^2$ of 0.7.

Previous studies report the mass of A746 as $M_{500,\text{SZ}} = 5.34^{+0.39}_{-0.40} \times 10^{14} M_\odot$ from the Planck SZ Survey (Planck Collaboration et al. 2016) and $M_{500,\text{X}} = 3.0 \pm 0.1 \times 10^{14} M_\odot$ from the X-ray temperature scaling relation (Rajpurohit et al. 2023). Our WL result gives $M_{500,\text{WL}} = 4.4 \pm 1.0 \times 10^{14} M_\odot$ at $R_{500}$, which is consistent with both the X-ray and SZ-based results.

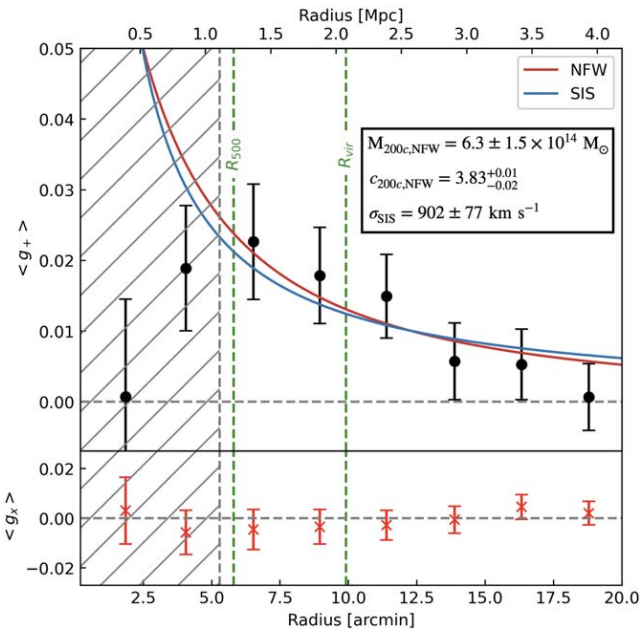

**Figure 5.** Reduced shear profile of A746. Top: Tangential shears are azimuthally averaged with respect to the central BCG. The displayed $1\sigma$ error bar accounts for both the ellipticity dispersion and measurement error. The red solid line indicates the best-fit NFW model, and the blue is the best-fit SIS model. The mass–concentration relation from Ishiyama et al. (2021) is assumed. The hatched region ($\lesssim R_{500}$) is ignored during the fit in order to minimize systematics (see the text). The green dashed line shows the virial radius from the best-fit NFW result. Bottom: The red crosses represent the cross-shear signal obtained by rotating galaxy images by 45°, which serves as a diagnostic for residual systematics. The cross shear is consistent with zero in this study.

## 6. Relic Power versus Cluster Mass

Studies of a radio relic population (e.g., de Gasperin et al. 2014; Nuza et al. 2017; Jones et al. 2023) show a significant correlation between the relic luminosity and the host cluster mass. This is perhaps because the total energy budget of the merger shock is related to the binding energy of clusters while the details are modulated by the energy dissipation rate, magnetic field, and acceleration efficiency (Poole et al. 2006; Hoeft et al. 2008; Pfrommer 2008; Skillman et al. 2011). Comparisons with theories have been performed. However, due to the small sample size, observational scatters remain still large (e.g., de Gasperin et al. 2014; Jones et al. 2023).

Figure 6 shows the scaling relation between the radio power of relics and cluster mass from de Gasperin et al. (2014) and Botteon et al. (2022) in comparison with the updated mass of A746. We adopted the monochromatic radio powers of the radio relics at 150 MHz from Rajpurohit et al. (2023) where $P_{150\,\mathrm{MHz,NW}} = 7.7 \times 10^{25}$ W Hz$^{-1}$, $P_{150\,\mathrm{MHz,R1}} = 0.38 \times 10^{25}$ W Hz$^{-1}$, and $P_{150\,\mathrm{MHz,R3}} = 1.1 \times 10^{25}$ W Hz$^{-1}$ for the NW, R1, and R3 relics, respectively. At the A746 mass ($M_{500,\mathrm{WL}} = 4.4 \pm 1.0 \times 10^{14}\,M_\odot$), the R1 and R3 luminosities are highly consistent with the literature results adopted here. The NW relic power is an order of magnitude higher than the best-fit literature relations. However, given the mass uncertainty, the tension is not significant. Nevertheless, the large differences in radio power among the three relics suggest that they may have originated from different mergers.

## 7. Discussion of the Merger Scenario

The multiwavelength data of A746 reveal several post-merger signatures including radio relics, X-ray shocks, and a disturbed ICM morphology. The presence of the symmetric double radio relics (NW relic and R3) suggests that a near head-on collision might have occurred along the east–west direction. The W and S substructures, defined by the galaxy overdensity, approximately align with the hypothesized merger axis (right panel of Figure 4). One challenge in this scenario is that the S subcluster is in close proximity to the smaller relic (R3) while the W subcluster is positioned midway between the two relics. Typically, clusters with symmetric double radio relics have a bimodal galaxy (and dark matter) distribution, and the two subclusters are positioned at approximately equal distances from their adjacent relics (e.g., Jee et al. 2015; Finner et al. 2021; Kim et al. 2021; Cho et al. 2022; Lee et al. 2022).

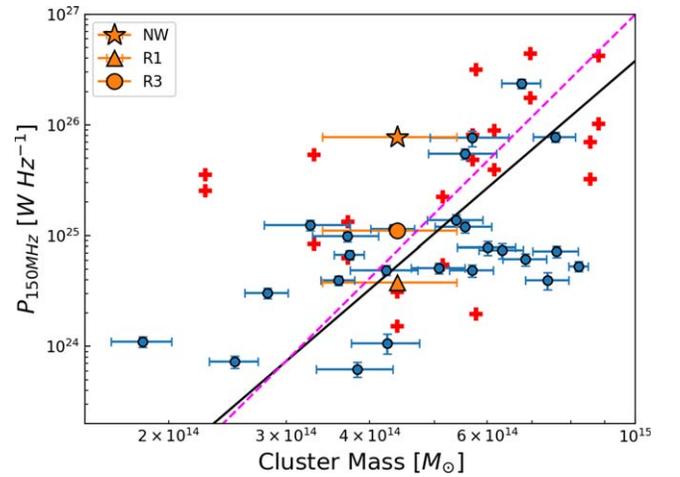

**Figure 6.** Scaling relation between the radio power of relics and the cluster mass. The orange markers indicate the radio relics in A746 with the WL mass. The red crosses and blue dots are radio relic samples adopted from de Gasperin et al. (2014) and Botteon et al. (2022), respectively. The magenta and black lines are the corresponding fits adopted from de Gasperin et al. (2014) and Jones et al. (2023) (candidate relics are excluded), respectively.

The deviation from this pattern in A746 suggests that possibly another merger involving a third subcluster occurred shortly after the initial binary merger. If we assume that the N substructure is the remnant of the third subcluster responsible for this hypothesized disruption, its trajectory could have been approximately from south to north. Interestingly, both the position and orientation of the R1 relic are consistent with the hypothesized merger axis of the second collision. It is possible that the N subcluster gravitationally interacted with both the S and W subclusters and affected their trajectories.

The V-shaped X-ray emission is reminiscent of the wake features observed in the El Gordo cluster (Menanteau et al. 2012). Although it is difficult to directly associate the orientations of the wake features with the aforementioned merger axes, A746's complex X-ray morphology with no distinct core is suggestive of multiple mergers.

We clarify that in the aforementioned merger scenario, we exclude the R2 radio emission, which is in the eastern periphery of A746. If we consider R2 as a relic connected to R1,[16] we suggest that perhaps R2 is generated during the north–south merger, which is also responsible for the R1 relic. To reconcile such a challenging scenario, a future study with deeper X-ray and radio data is required.

---

[16] However, note that R1 and R2 are viewed as having different origins in Rajpurohit et al. (2023).

## 8. Conclusions

A746 is a highly disturbed cluster possessing prominent merging features including double radio relics, two isolated radio relics, disturbed X-ray morphology, and complex galaxy/mass substructures, challenging our understanding of their origins. We have presented a WL analysis using Subaru/Hyper Suprime-Cam observations and have re-examined the cluster's galaxy distribution with MMT/Hectospec spectroscopic data. Our investigation of the line-of-sight velocity distribution and the correlation with the spatial distribution suggests that perhaps the mergers might be happening in the plane of the sky or its merging subclusters have already reached their apocenters. The mass distribution is characterized by the main peak coinciding with the geometric center of the disturbed X-ray emission. From this main peak, we have identified two elongations toward the north and west. These features approximately follow the cluster galaxy distribution. Although A746 is a complex merger, the A746 global mass estimated by WL analysis is in good agreement with the previous results obtained with SZ and X-ray observations.

Based on the identification of three substructures in both galaxy and mass distributions, as well as the presence of double radio relics and the northern relic, we propose a merging scenario involving two successive mergers with three subclusters. In our scenario, the first near head-on collision happened along the east–west direction between two subclusters, producing the current double radio relics. Shortly after this merger, the third subcluster passed from south to north and affected the trajectories of the first two subclusters, giving rise to the northern radio relic. Nevertheless, we conclude that detailed numerical simulations and deeper observations are required to comprehend the complex distributions of mass, galaxies, ICM, and radio features.

## Acknowledgments

M.J.J. acknowledges support for the current research from the National Research Foundation (NRF) of Korea under the programs 2022R1A2C1003130 and RS-2023-00219959. This work was supported by K-GMT Science Program (PID: GEMINI-KR-2022B-011 and MMT-2019A-004) of Korea Astronomy and Space Science Institute (KASI). D.W. was supported by NSF award 2308383. A.S. acknowledges the support of a Clay Fellowship. R.J.v.W acknowledges support from the ERC Starting Grant ClusterWeb 804208. P.D. acknowledges the Future Faculty Leaders Fellowship at the Center for Astrophysics, Harvard-Smithsonian. A.B. acknowledges financial support from the European Union—Next Generation EU. L.L. acknowledges the financial contribution from INAF grant 1.05.12.04.01. This research is based on data collected at the Subaru Telescope (PID: S22B-TE011-GQ), via the time exchange program between Subaru and the international Gemini Observatory, a program of NSF's NOIRLab. The Subaru Telescope is operated by the National Astronomical Observatory of Japan. We are honored and grateful for the opportunity to observe the Universe from Maunakea, which has cultural, historical, and natural significance in Hawaii. This paper makes use of LSST Science Pipelines software developed by the Vera C. Rubin Observatory. We thank the Rubin Observatory for making their code available as free software at https://pipelines.lsst.io. Observations reported here were obtained at the MMT Observatory, a joint facility of the Smithsonian Institution and the University of Arizona.

*Facilities:* LOFAR, MMT (Hectospec), Subaru (Hyper Suprime-Cam), XMM

## Appendix A
## Star Subtraction

We aim to minimize the effect of the bright star f UMa ($V = 4.48$ mag) on our source detection and shape analysis by subtracting its light profile. The radial profile does not follow a simple analytic model because of the saturation at the core. Nevertheless, as described in Section 2, it ensures an axisymmetric distribution. To extract the radial profile of the star, we set up radial bins centered at the star. Then, for each radial bin, we measured a $2\sigma$-clipped mean. In this procedure, we masked out other astronomical sources using the segmentation map output by SExtractor. We iterated this radial profile extraction for each exposure and subtracted the resulting radial profile before the final co-addition.

Figure A1 shows the star-subtracted image in comparison with the original mosaic image. Although far from perfect, optical ghosts around the star are substantially reduced in the $r2$ band. However, the central region ($R < 160''$) of the star remains unusable for our scientific study, and thus was masked out in the subsequent analysis.

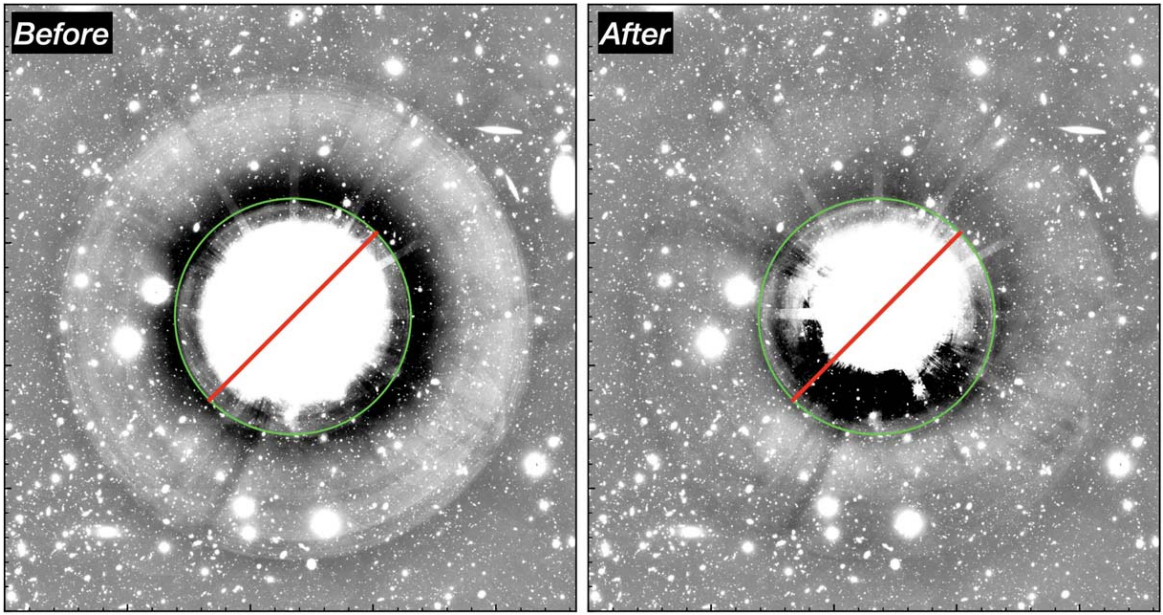

**Figure A1.** Image restoration near the very luminous star f UMa. The concentric optical ghosts are reduced and the over-subtraction is mitigated. The green circle indicates the masked region during the analysis. The two (before and after) images are displayed using the same dynamic scale.

## Appendix B
## Spectral Decomposition of BCG-S

Our visual inspection of BCG-S suggests that it might be a superposition of two galaxies. Our spectral decomposition shows that indeed the BCG-S spectrum is a mixture of two galaxies: one at the cluster redshift and the other at a slightly higher redshift. Below we provide details.

The spectroscopic redshift of BCG-S from the SDSS archive is $z = 0.2888$. Its color is bluer than the cluster red sequence by ∼0.3 (red cross in Figure 3). Taken at face value, this would be a remarkable background galaxy—a bluish galaxy more luminous than any cluster galaxy yet with no companions at its redshift. However, close inspection of the imaging reveals overlapping objects near the core. The objects are not properly deblended since the peak-to-peak contrast is low and they are clustered within a compact region. The combination of a luminous red cluster galaxy and a star-forming galaxy would yield a blend with the observed photometric properties. The SDSS spectroscopic fibers gather light from a 3″ diameter region. Thus, any such blending would also affect the spectrum.

We therefore looked more closely at the SDSS spectrum (Figure B1 with data in blue). The SDSS model (orange) clearly fits the emission lines at $z = 0.2888$ but is not a good fit to the continuum. The model overpredicts the continuum flux at some wavelengths and underpredicts at others; in particular, the data show an apparent 4000 Å break at an observed wavelength of about 4800 Å, which for a single-galaxy model is inconsistent with the [O II] 3727 line observed at nearly the same wavelength.

We modeled the spectrum using two SDSS templates at different redshifts: a star-forming galaxy at $z = 0.2888$ and an early-type galaxy with redshift as a free parameter. The flux ratio between the two galaxies was a second free parameter, while the summed model flux was set to match the observed flux. We found $z = 0.21339 \pm 0.00004$ for the early-type component. The blended model (shown in gray in Figure B1) lowers $\chi^2$ by 140, sufficient to justify the additional parameters. In addition, the insets in Figure B1 demonstrate how well the blended model matches absorption features not present in the star-forming galaxy model that describes the emission lines. In addition to the insets shown, we note that Mg 5167 (observed at 6270 Å) and H$\beta$ (observed at 5900 Å) absorption are well modeled. Furthermore, the continuum is better fit in many areas. Finally, the amplitudes of the emission lines are better fit because the amplitude of the star-forming galaxy is reduced relative to the SDSS model.

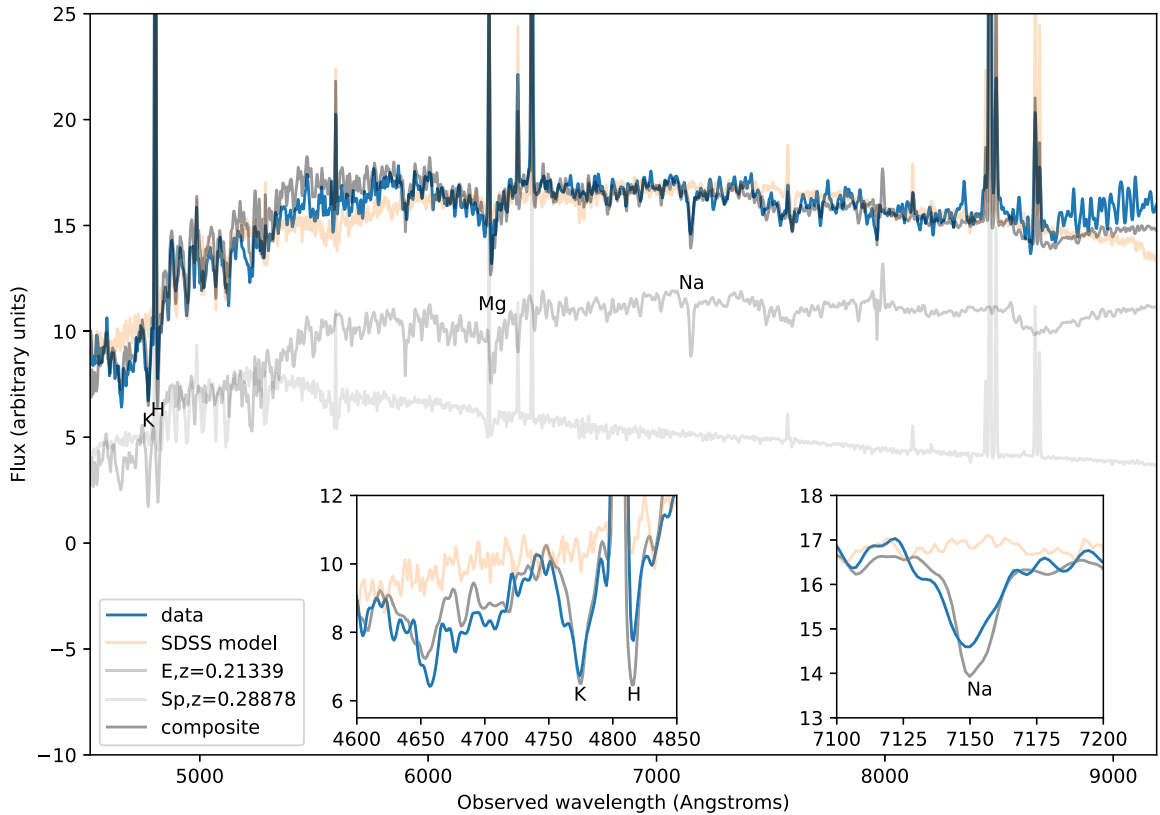

**Figure B1.** Spectroscopic decomposition of BCG-S. The blue solid line shows the observed spectrum and the orange line represents the SDSS model ($z = 0.2888$). The spectrum is a much better fit by adding an early-type galaxy; with the redshift of that galaxy as a free parameter, we find $z = 0.21339 \pm 0.00004$. The composite model is shown in dark gray (which appears black when plotted over the data) and its components are plotted in lighter shades of gray. The insets show some of the absorption features that are well fit with the composite model but not the single-galaxy SDSS model. This demonstrates that BCG-S is indeed in the cluster, and is blended with a background galaxy that contributes blue light and emission lines.

## ORCID iDs

K. HyeongHan ⓘ https://orcid.org/0000-0002-2550-5545
H. Cho ⓘ https://orcid.org/0000-0001-5966-5072
M. James Jee ⓘ https://orcid.org/0000-0002-5751-3697
D. Wittman ⓘ https://orcid.org/0000-0002-0813-5888
S. Cha ⓘ https://orcid.org/0000-0001-7148-6915
W. Lee ⓘ https://orcid.org/0000-0002-1566-5094
K. Finner ⓘ https://orcid.org/0000-0002-4462-0709
K. Rajpurohit ⓘ https://orcid.org/0000-0001-7509-2972
M. Brüggen ⓘ https://orcid.org/0000-0002-3369-7735
W. Forman ⓘ https://orcid.org/0000-0002-9478-1682
C. Jones ⓘ https://orcid.org/0000-0003-2206-4243
R. van Weeren ⓘ https://orcid.org/0000-0002-0587-1660
A. Botteon ⓘ https://orcid.org/0000-0002-9325-1567
L. Lovisari ⓘ https://orcid.org/0000-0002-3754-2415
A. Stroe ⓘ https://orcid.org/0000-0001-8322-4162
P. Domínguez-Fernández ⓘ https://orcid.org/0000-0001-7058-8418
E. O'Sullivan ⓘ https://orcid.org/0000-0002-5671-6900
J. Vrtilek ⓘ https://orcid.org/0009-0007-0318-2814

## References

Ahumada, R., Prieto, C. A., Almeida, A., et al. 2020, ApJS, 249, 3
Aihara, H., AlSayyad, Y., Ando, M., et al. 2022, PASJ, 74, 247
Almeida, A., Anderson, S. F., Argudo-Fernández, M., et al. 2023, ApJS, 267, 44
Bertin, E., & Arnouts, S. 1996, A&AS, 117, 393
Bertin, E., Mellier, Y., Radovich, M., et al. 2002, in ASP Conf. Ser. 281, Astronomical Data Analysis Software and Systems XI, ed. D. A. Bohlender, D. Durand, & T. H. Handley (San Francisco, CA: ASP), 228
Bonafede, A., Giovannini, G., Feretti, L., Govoni, F., & Murgia, M. 2009, A&A, 494, 429
Bosch, J., AlSayyad, Y., Armstrong, R., et al. 2019, in ASP Conf. Ser. 523, Astronomical Data Analysis Software and Systems XXVII, ed. P. J. Teuben (San Francisco, CA: ASP), 521
Bosch, J., Armstrong, R., Bickerton, S., et al. 2018, PASJ, 70, S5
Botteon, A., Shimwell, T. W., Cassano, R., et al. 2022, A&A, 660, A78
Brunetti, G., & Jones, T. W. 2014, IJMPD, 23, 1430007
Cassano, R., Ettori, S., Giacintucci, S., et al. 2010, ApJL, 721, L82
Cha, S., & Jee, M. J. 2022, ApJ, 931, 127
Cho, H., James Jee, M., Smith, R., Finner, K., & Lee, W. 2022, ApJ, 925, 68
Clowe, D., Bradač, M., Gonzalez, A. H., et al. 2006, ApJL, 648, L109
Cuciti, V., Cassano, R., Brunetti, G., et al. 2015, A&A, 580, A97
Dawson, W. A., Jee, M. J., Stroe, A., et al. 2015, ApJ, 805, 143
Dawson, W. A., Schneider, M. D., Tyson, J. A., & Jee, M. J. 2016, ApJ, 816, 11
de Gasperin, F., van Weeren, R. J., Brüggen, M., et al. 2014, MNRAS, 444, 3130
Ducati, J. R., Bevilacqua, C. M., Rembold, S. B., & Ribeiro, D. 2001, ApJ, 558, 309
Fabricant, D., Fata, R., Roll, J., et al. 2005, PASP, 117, 1411
Finner, K., HyeongHan, K., Jee, M. J., et al. 2021, ApJ, 918, 72
Finner, K., James Jee, M., Webb, T., et al. 2020, ApJ, 893, 10
Finner, K., Jee, M. J., Golovich, N., et al. 2017, ApJ, 851, 46
Finner, K., Randall, S. W., Jee, M. J., et al. 2023, ApJ, 942, 23
Fischer, P., & Tyson, J. A. 1997, AJ, 114, 14
Furusawa, H., Koike, M., Takata, T., et al. 2018, PASJ, 70, S3
Golovich, N., Dawson, W. A., Wittman, D. M., et al. 2019a, ApJS, 240, 39
Golovich, N., Dawson, W. A., Wittman, D. M., et al. 2019b, ApJ, 882, 69
Golovich, N., van Weeren, R. J., Dawson, W. A., Jee, M. J., & Wittman, D. 2017, ApJ, 838, 110

Gruen, D., Seitz, S., & Bernstein, G. M. 2014, PASP, 126, 158
Guennou, L., Adami, C., Durret, F., et al. 2014, A&A, 561, A112
Hoeft, M., Brüggen, M., Yepes, G., Gottlöber, S., & Schwope, A. 2008, MNRAS, 391, 1511
HyeongHan, K., Jee, M. J., Cha, S., & Cho, H. 2023, NatAs, in press
HyeongHan, K., Jee, M. J., Rudnick, L., et al. 2020, ApJ, 900, 127
Ishiyama, T., Prada, F., Klypin, A. A., et al. 2021, MNRAS, 506, 4210
Jee, M. J., Blakeslee, J. P., Sirianni, M., et al. 2007, PASP, 119, 1403
Jee, M. J., Dawson, W. A., Stroe, A., et al. 2016, ApJ, 817, 179
Jee, M. J., Stroe, A., Dawson, W., et al. 2015, ApJ, 802, 46
Jee, M. J., & Tyson, J. A. 2009, ApJ, 691, 1337
Jee, M. J., & Tyson, J. A. 2011, PASP, 123, 596
Jee, M. J., Tyson, J. A., Schneider, M. D., et al. 2013, ApJ, 765, 74
Jones, A., de Gasperin, F., Cuciti, V., et al. 2023, A&A, 680, A31
Kaiser, N., & Squires, G. 1993, ApJ, 404, 441
Kawanomoto, S., Uraguchi, F., Komiyama, Y., et al. 2018, PASJ, 70, 66
Kim, J., Jee, M. J., Hughes, J. P., et al. 2021, ApJ, 923, 101
Komiyama, Y., Obuchi, Y., Nakaya, H., et al. 2018, PASJ, 70, S2
Kurtz, M. J., & Mink, D. J. 1998, PASP, 110, 934
Lee, W., Cha, S., Jee, M. J., et al. 2023, ApJ, 945, 71
Lee, W., James Jee, M., Finner, K., et al. 2022, ApJ, 924, 18
Mandelbaum, R. 2018, ARA&A, 56, 393
Markevitch, M., Sarazin, C. L., & Vikhlinin, A. 1999, ApJ, 521, 526
Menanteau, F., Hughes, J. P., Sifón, C., et al. 2012, ApJ, 748, 7
Miyazaki, S., Komiyama, Y., Kawanomoto, S., et al. 2018, PASJ, 70, S1
Navarro, J. F., Frenk, C. S., & White, S. D. M. 1996, ApJ, 462, 563
Navarro, J. F., Frenk, C. S., & White, S. D. M. 1997, ApJ, 490, 493
Nuza, S. E., Gelszinnis, J., Hoeft, M., & Yepes, G. 2017, MNRAS, 470, 240
Okabe, N., Bourdin, H., Mazzotta, P., & Maurogordato, S. 2011, ApJ, 741, 116
Pfrommer, C. 2008, MNRAS, 385, 1242
Planck Collaboration, Ade, P. A. R., Aghanim, N., et al. 2016, A&A, 594, A27
Poole, G. B., Fardal, M. A., Babul, A., et al. 2006, MNRAS, 373, 881
Ragozzine, B., Clowe, D., Markevitch, M., Gonzalez, A. H., & Bradač, M. 2012, ApJ, 744, 94
Rajpurohit, K., Lovisari, L., Botteon, A., et al. 2023, arXiv:2309.01716
Ricker, P. M., & Sarazin, C. L. 2001, ApJ, 561, 621
Rowe, B. T. P., Jarvis, M., Mandelbaum, R., et al. 2015, A&C, 10, 121
Schneider, P. 2005, arXiv:astro-ph/0509252
Seitz, C., & Schneider, P. 1997, A&A, 318, 687
Shupe, D. L., Laher, R. R., Storrie-Lombardi, L., et al. 2012, Proc. SPIE, 8451, 84511M
Skillman, S. W., Hallman, E. J., O'Shea, B. W., et al. 2011, ApJ, 735, 96
Tonry, J., & Davis, M. 1979, AJ, 84, 1511
van Weeren, R. J., Brüggen, M., Röttgering, H. J. A., et al. 2011, A&A, 533, A35
van Weeren, R. J., de Gasperin, F., Akamatsu, H., et al. 2019, SSRv, 215, 16
van Weeren, R. J., Röttgering, H. J. A., Brüggen, M., & Hoeft, M. 2010, Sci, 330, 347
Weaver, J. R., Kauffmann, O. B., Ilbert, O., et al. 2022, ApJS, 258, 11
Wittman, D., Dell'Antonio, I. P., Hughes, J. P., et al. 2006, ApJ, 643, 128